\documentclass[aps,twocolumn,showpacs,superscriptaddress,floats,prb]{revtex4}
\usepackage[german,english]{babel}
\usepackage{amsmath,amssymb}
\usepackage{graphicx}

\newcommand{\mr}[1]{{{\mathrm{#1}}}}

\newcommand{\gz}{g}

\begin{document}

\title{
Boundary quantum criticality \\
in models of magnetic impurities coupled to bosonic baths
}

\author{Serge Florens}
\affiliation{
Institut N\'eel, CNRS and Universit\'e
Joseph Fourier, BP 166, 38042 Grenoble, France
}
\author{Lars Fritz}
\author{Matthias Vojta}
\affiliation{
Institut f\"{u}r Theoretische Physik, Universit\"{a}t zu K\"{o}ln,
Z\"ulpicher Str. 77, 50937 K\"{o}ln, Germany
}

\date{March 23, 2007}

\begin{abstract}
We investigate quantum impurity problems, where a local magnetic moment
is coupled to the spin density of a bosonic environment,
leading to bosonic versions of the standard Kondo and Anderson impurity models.
In a physical situation, these bosonic environments can correspond either to
deconfined spinons in certain classes of Z$_2$ frustrated antiferromagnets,
or to particles in a multicomponent Bose gase
(in which case the spin degree of freedom is attributed to hyperfine levels).
Using renormalization group techniques, we establish that our impurity models,
which feature an exchange interaction analogous to Kondo impurities in Fermi liquids,
allow the flow towards a stable strong-coupling state.
Since the low-energy bosons live around a single point in momentum space,
and there is no Fermi surface,
an impurity quantum phase transition occurs at intermediate coupling,
separating screened and unscreened phases.
This behavior is qualitatively different from previously studied spin-isotropic
variants of the spin-boson model, which display stable intermediate-coupling
fixed points and no screening.
\end{abstract}

\maketitle


\section{Introduction}

The investigation of quantum phase transitions in strongly correlated electronic
systems has been an active field of research during the last two decades.\cite{Sachdevbook}
Recently, there has been growing interest in so-called impurity quantum phase
transitions --
these correspond to qualitative changes in the ground-state properties of
discrete degrees of freedom (a spin $1/2$ in the simplest case)
upon changing their couplings to an external environment.
On the one hand, impurity quantum phase transitions can be realized, e.g., in mesoscopic systems,
such as quantum dot devices, which provide the high tunability required to
access the transition point.
On the other hand, the study of impurity problems is motivated through the dynamical mean-field
theory\cite{Metzner,Georges} (DMFT),
which maps lattice models of strongly correlated electrons --
assuming a momentum-independent self energy -- onto effective
self-consistent impurity models.
In this language, certain bulk quantum phase transition correspond
to impurity transitions in Anderson or Kondo-type models.

Impurity quantum phase transition occur in different contexts and setups.
One possibility is a single magnetic impurity coupled
to an unconventional fermionic host.
A paradigmatic example being the so-called
pseudogap Kondo model,\cite{Withoff} originally motivated by the study of
magnetic impurities in $d$-wave superconductors,
where a spin $1/2$ is coupled to the spin density of
fermionic quasiparticles with a power-law density of states (DOS),
$\rho(\omega) \propto |\omega|^r$.
The lack of low-energy states (as compared to the standard Kondo model\cite{Hewson})
causes a competition between the local-moment dynamics and the formation of
a strong-coupling ground state and leads to rich quantum critical
behavior.\cite{Withoff,GBI,Fritz,Fritz1}
Other fermionic impurity models display quantum phase transitions due to
the competition between differently screened states, with the
two-channel Kondo\cite{Cox,Potok} and the two-impurity Kondo
models\cite{Jones,ALJ,Zarand} being popular examples.
The majority of these models were recently reviewed in Ref.~\onlinecite{Matthiasreview}.

Quantum impurities with a {\it bosonic} environment constitute a distinct class of models
which can feature quantum phase transitions as well.
A well-studied model is the spin-boson model, originally introduced to capture
dissipation effects at the quantum level. Here, a generic two-level system (dubbed spin)
is subject to the competing influence of a transverse field (causing tunneling between
the two levels) and a longitudinally-coupled set of harmonic oscillators (providing
friction).
In the case of ohmic dissipation, this results in a Kosterlitz-Thouless phase transition
at zero temperature, tuned by a variation of the dissipation strength.\cite{Zwerger}
Note that the ohmic spin-boson model is equivalent to a Kondo model with
anisotropic spin exchange terms, and hence does not constitute a distinct
universality class.
(This is different for sub-ohmic damping, see Refs.~\onlinecite{VTB,Matthiasreview}.)
The topic of impurities in quantum magnets has motivated studies of the
so-called Bose-Kondo model
(which may also be viewed as a spin-isotropic spin-boson model with zero transverse
field) -- this model has a a stable intermediate-coupling fixed point
and no screening;\cite{vojta0,vojta,ssmv03,troyer,sandvik}
this will be reviewed briefly in Sec.~\ref{vojtabura}.
We note that the above-mentioned bosonic impurity models can be extended
to include a fermionic environment as well, resulting in Bose-Fermi Kondo Hamiltonians,
where the interactions of the impurity with the bosonic and fermionic baths
compete.\cite{Si,Sengupta,Zhu,Zarand2,MVMK}
Models of this type have been proposed to describe quantum criticality
in heavy-fermion compounds within extended DMFT,\cite{Si}
and could as well be realized in certain quantum-dot setups.\cite{LeHur}

We have recently introduced\cite{Florens} a distinct class of
SU(2)-symmetric bosonic impurity models,
where the environment consists of spin-1/2 objects, i.e., spinons constituting
the elementary excitations of certain frustrated quantum magnets.
In contrast to the spin-boson type models mentioned above, where
the environment always plays the role of a fluctuating (classical) magnetic field,
our models display true quantum dynamics and allow singlet formation
accompanied by complete screening.

The purpose of this paper is to give a comprehensive renormalization group (RG) analysis
of various versions of these bosonic Kondo models, thus providing details
which were left out in the short report of Ref.~\onlinecite{Florens}:
(i) The first example concerns the elementary bosonic excitations (spinons) in the Z$_2$
spin-liquid state of a triangular Heisenberg model.\cite{Park}.
This model will be shown shown to display an impurity quantum phase transition,
bearing some resemblance to the fermionic pseudogap Kondo model.
(ii) A second example that follows the same philosophy concerns the setup of a ``magnetic
impurity'' embedded in an optical lattice filled with a multicomponent bosonic
gas.\cite{Zoller,williams}
Here, the bosons are not collective modes originating from an underlying magnetic state,
but are rather real bosonic atoms.
In both cases, we derive the relevant quantum-field theory and study it
using RG combined with epsilon expansion techniques. Although the structure of the
theories is reminiscent of fermionic Kondo models, there are important differences:
we demonstrate the importance of potential-scattering effects at the impurity location
as well as the influence of bulk interactions.

Note that this paper will only study the RG flow at weak coupling;
for situations with runaway flow to strong coupling -- indicative of true screening --
different techniques need to be applied.
We have demonstrated this in Ref.~\onlinecite{Florens}, using
strong-coupling toy models and large-$N$ techniques; this will not be repeated here.
Furthermore, we restrict our attention to situations where the bulk state does not
break a local symmetry, i.e. we focus on paramagnetic phases including the
quantum critical points.

The plan of the paper is as follows.
In Sec.~\ref{orderparameters} we summarize the order-parameter description
of nearly-critical antiferromagnets, distinguishing collinear and non-collinear ordering.
In the non-collinear case of interest to us, deconfined bosonic spinons\cite{Park}
emerge which correspond to fractional spin $1/2$ excitations in a Z$_2$ spin liquid
phase.\cite{chubukov,isakov,azaria}
In Sec.~\ref{vojtabura} we review the results of earlier work\cite{vojta0,vojta,ssmv03} on
a magnetic moment in a collinear antiferromagnet, described by
a conventional $\phi^4$ theory with triplon excitations,
leading to a spin-isotropic spin-boson-like model without screening.
Sec.~\ref{RG} is devoted to the central topic of this paper,
namely the physics of a magnetic impurity in a Z$_2$ spin liquid.
We present a microscopic derivation of the impurity action,
which we analyze by RG methods,
extending our earlier calculation\cite{Florens} to two-loop level.
Furthermore we propose a way to properly incorporate the bulk interactions
using a mapping to an effective bosonic Anderson impurity model.
Finally, in Sec.~\ref{canonical} we study how an impurity two-level system couples
to an atomic gas of canonical bosons carrying an internal degree of freedom
(pseudospin).
Again, we determine the phase diagram and critical properties using RG.
A brief outlook will close the paper.
Details of the RG calculations are deferred to the appendices.
There, we also point out that the RG flow derived in Sec.~\ref{canonical}
is identical to the one of a fully asymmetric fermionic Kondo model, which
may arise in the DMFT context.


\section{Field theories of nearly-critical quantum antiferromagnets}
\label{orderparameters}

Different types of magnetic Mott insulators naturally call for different
types of order-parameter theories to describe the physics near the transition
towards the symmetry-broken phase.
In general, introducing an order parameter $\phi$ for the magnetic fluctuations
near the ordering wavevector $\vec K$, a gradient expansion of the action
naturally leads to a standard multicomponent $\phi^4$ theory
(or a non-linear sigma model).
However, care is required for non-collinear spin correlations, as
shown below.

Assuming a magnetic order characterized by a single ordering
wave vector $\vec{K}$, the local spin operators can be parametrized as
\begin{eqnarray}
\vec{S}_i = \mathfrak{Re} \left(\vec\phi e^{i\vec{K}\cdot \vec{x}_i}\right) =
\vec{n}_1 \cos(\vec{K}\cdot \vec{x}_i)+
\vec{n}_2 \sin(\vec{K}\cdot \vec{x}_i) \; ,
\label{equ:orderparameter}
\end{eqnarray}
where $\vec{x}_i$ denotes the position of lattice site $i$, and
$\vec\phi(\vec x,\tau) = \vec{n}_1+i\vec{n}_2$ is a complex vector field in spin space
which describes the order-parameter direction and varies slowly in space and time.

\subsection{Collinear magnetic order}

For a state with static collinear order we have
fixed vectors $\vec{n}_{1,2}$ with $\vec{n}_1 \times \vec{n}_2=0$;
the $\langle\vec{S}_i \rangle$ in Eq.~\eqref{equ:orderparameter} on all
sites $i$ point parallel or antiparallel w.r.t. a common axis.
The undoped insulator La$_2$CuO$_4$ is of this type
(if we neglect the small spin canting due to Dzyaloshinski-Moriya interactions),
with in-plane ordering wave-vector $\vec{K}=(\pi,\pi)$
(the lattice spacing is unity here and in the following).
For this specific value of $\vec{K}$ on a square lattice
the order parameter is independent of $\vec{n}_2$.

Near quantum criticality,
one is lead to the usual O(3) quantum $\phi^4$ theory in imaginary time:
\begin{equation}
\label{Scoll}
\mathcal{S}_b = \int d^d x d \tau \left[
|\partial_x \vec\phi|^2 + |\partial_\tau \vec\phi|^2 + g_0 (\vec\phi^2)^2
\right] \,.
\end{equation}
For $\vec{K}=(\pi,\pi)$ the vector field $\vec\phi$ is now {\em real};
for commensurate $\vec{K}$, $\vec\phi$ is complex, but the phase
degree of freedom is gapped.
This theory has been studied at length in the literature,\cite{Sachdevbook,Park}
and its critical properties are relevant for the quantum phase transition
from a N\'eel state to a gapped paramagnet in several classes
of unfrustrated antiferromagnets such as the bilayer square-lattice Heisenberg model and
coupled spin-ladder models.

For general incommensurate $\vec{K}$, one has to parametrize
$\vec\phi = \vec n e^{i\Theta}$, with a real vector $\vec n$ and a phase field $\Theta$.
Interesting critical behavior may obtain here\cite{ssmorinari}, which
is not in the focus of this paper.

\subsection{Non-collinear magnetic order}
\label{noncollinearorder}
Non-collinear order is present if $\vec{n}_1 \times \vec{n}_2 \neq 0$ in
\eqref{equ:orderparameter}; the order may be coplanar or fully three-dimensional.
The former situation is e.g. realized in triangular quantum antiferromagnets,
where the $\langle\vec{S}_i \rangle$ now lie in a plane in spin space,
rather than along a single axis.

The simplest case of a non-collinear ordered state is the one where
\begin{eqnarray}
\vec{n}_1 \cdot \vec{n}_2=0 \quad ; \quad \vec{n}_1^2=\vec{n}_2^2 = 1,
\label{spiral}
\end{eqnarray}
corresponding to a spin spiral.
An elegant way to resolve the constraints is via a parametrization
of the order parameter according to\cite{chubukov}
\begin{eqnarray}
\vec\phi = \vec{n}_1+i \vec{n}_2= \epsilon_{\sigma \sigma'} z^{\phantom{*}}_{\sigma'}
\frac{\vec{\sigma}_{\sigma \tau}}{2} z^{\phantom{\dagger}}_\tau \; ,
\label{couplingform}
\end{eqnarray}
with the single constraint $\sum_\sigma|z^{\phantom{*}}_\sigma|^2=1$ (here
$\sigma=\uparrow,\downarrow$),
$\vec{\sigma}$ is the vector of Pauli matrices, and
$\epsilon_{\sigma \sigma'}$ is the fully antisymmetric matrix.
The objects $z_\sigma$ represent complex numbers
(and hence are not canonical bosons), they describe a mapping from the
four-dimensional unit sphere S$_3$ to S$_2$.
Obviously, this representation is unique up to a local Z$_2$ transformation
$z_\sigma \to \eta z_\sigma$ with $\eta= \pm 1$.

Thus, in the non-collinear case described by (\ref{spiral},\ref{couplingform}),
the long-distance fluctuations near a magnetic quantum critical point
are described by a quantum field theory in terms of bosonic
spinons:\cite{chubukov}
\begin{eqnarray}
\mathcal{S}&=& \frac{1}{\gz} \int d^dx  d \tau
\left ( |\partial_x z_\sigma|^2+|\partial_\tau z_\sigma|^2 \right ) \nonumber \\
&+& \int d^dx d\tau \, \kappa(x,\tau) \Big(\sum_\sigma |z_\sigma|^2-1 \Big)
\label{equ:spinonaction}
\end{eqnarray}
where the integration over the auxiliary field $\kappa$ enforces the
constraint on $\sum_\sigma|z_\sigma|^2$.
In Eq.~\eqref{equ:spinonaction}
an additional Z$_2$ gauge field, corresponding to the above-mentioned
gauge redundancy, has been omitted -- this is justified provided that the
gauge field is in a deconfined phase with massive fluctuations,
which is the case for the triangular antiferromagnet.
Note that the paramagnetic phase of \eqref{equ:spinonaction} is a
Z$_2$ spin liquid with gapped, deconfined bosonic
spinons.\cite{chubukov,isakov,azaria}

Let us note here that the fluctuations in a {\it collinear} magnet
may also be written in terms of bosonic spinons, leading to
a Schwinger boson representation:\cite{Auerbach,Fradkin}
\begin{eqnarray}
\vec{S} (\vec{x}) = \sum_{\sigma,\sigma'}z^*_\sigma (\vec{x})
\frac{\vec{\sigma}_{\sigma \sigma'}}{2}z^{\phantom{*}}_{\sigma'} (\vec{x})\; ,
\end{eqnarray}
with the constraint $\sum_\sigma |z^{\phantom{*}}_\sigma|^2=1$.
Note the difference with Eq.~\eqref{couplingform}:
in contrast to the non-collinear case, the accompanying gauge redundancy
here is of U(1) symmetry.
In such a situation, the gauge-field fluctuations are massless
and cannot be ignored in a proper description of the bulk magnetism.
The associated compact gauge theory is in general confining,\cite{Shenker}
implying that bosonic spinons are bound in pairs
and leading to the standard $\phi^4$ theory \eqref{Scoll} for triplon
fluctuations.\cite{ChubuStar}

U(1) spin liquids with deconfined spinons may nevertheless arise:
a special situation here are so-called deconfined critical
points,\cite{deconfined} where a description in terms of bosonic spinons
coupled to a non-compact U(1) gauge field applies
(which implies the suppression of hedgehogs\cite{Motrunich}).
Impurities characterized by a gauge charge and embedded in a U(1) spin liquid
have been studied in Ref.~\onlinecite{Kolezhuk};
the interplay with a magnetic impurity moment would be interesting to study,
but is beyond the scope of the present paper.


\section{A magnetic impurity in a quantum antiferromagnet with collinear order}
\label{vojtabura}

Magnetic impurities embedded in a nearly critical collinear quantum
antiferromagnet were analyzed in detail in Refs.~\onlinecite{vojta0,vojta,ssmv03,troyer,sandvik}.
The bulk action ${\cal S}_b$ (\ref{Scoll}),
describing the physics close to the transition between the ordered N\'eel state and
the disordered spin-gap state, is supplemented by the
coupling to an impurity quantum spin $1/2$ $\vec{S}$,
leading to:
\begin{eqnarray}
\mathcal{S} & = & \mathcal{S}_b + \mathcal{S}_ {\textrm{imp}}
+ \mathcal{S}_{\textrm{Berry}}[\vec{S}]\,,\nonumber\\
\mathcal{S}_{\textrm{imp}}&=& \gamma_0 \int d \tau
\vec S(\tau) \cdot \vec\phi (\vec{x}=0,\tau)
\label{fluctfield}
\end{eqnarray}
where $\vec\phi(\vec{x},\tau)$ is the real order-parameter field of (\ref{Scoll}),
and $\mathcal{S}_{\textrm{Berry}}[\vec{S}]$ encodes the dynamics of the
impurity spin $\vec{S}$.
A weak-coupling RG analysis\cite{vojta0,vojta}
leads to the RG beta function for the dimensionless impurity coupling $\gamma$:
\begin{eqnarray}
\label{betagamma}
\beta (\gamma)=
\frac{d\gamma}{d\ln\mu} =
-\frac{\epsilon \gamma}{2}+\gamma^3+\frac{5g^2\gamma}{144}
\end{eqnarray}
where $\epsilon=3-d$, $\mu$ is a renormalization scale, and the sign of
the beta function is such that negative terms are RG relevant in the infrared.
In Eq.~(\ref{betagamma}),
$g$ denotes the renormalized bulk coupling in $\mathcal{S}_b$ (\ref{Scoll})
which flows as in the standard $\phi^4$ theory.\cite{Zinn}
From Eq.~(\ref{betagamma}) one finds a fixed-point value of the impurity
coupling as $\gamma^{\ast 2} = \epsilon/2 + \mathcal{O}(\epsilon^2)$.
The critical fluctuations of the bulk plus impurity problem are thus controlled
by a fixed point with couplings $g^*$ (bulk) and $\gamma^{\ast 2}$ (impurity), which are
both of order $\epsilon$, i.e., perturbatively accessible near $d=3$ bulk dimensions.

Most importantly, the flow of $\gamma$ (\ref{betagamma}) near $\gamma^*$ is infrared stable (!),
in sharp contrast to (for example) the impurity quantum critical point of the
pseudogap Kondo model.\cite{Withoff, Fritz, Fritz1}
The absence of runaway flow in Eq.~(\ref{betagamma}) suggests
the absence of a strong-coupling phase with complete screening,
even at the bulk critical point with gapless bosonic modes.
In a sense, this is not unexpected, as Eq.~(\ref{fluctfield}) describes the classical
fluctuations of an effective magnetic field coupled to the impurity, being
unable to quench the entropy.

As a consequence of \eqref{betagamma}, no quantum phase transitions occur in
the model (\ref{fluctfield}).
The impurity properties are controlled by a stable intermediate-coupling fixed
point, which describes a non-trivially fluctuating fractional-spin
state.
Near this fixed point, the impurity susceptibility and entropy obey\cite{vojta}
\begin{eqnarray}
\lim_{T\to 0} T \chi_{\rm imp} = {\cal C}_1(d) ,~
\lim_{T\to 0} S_{\rm imp} = {\cal C}_2(d) \,.
\label{fracspin}
\end{eqnarray}
with universal constants ${\cal C}_1(d)$, ${\cal C}_2(d)$
(which depend on dimensionality $d$ and impurity spin size $S$ only).
We note that the problem of an impurity in a collinear magnet can
equivalently be formulated using a non-linear sigma model for the bulk\cite{ssmv03} --
this leads to an expansion in $\epsilon = d-1$, with
similar impurity properties.
Thus, the fractional-spin state is present for all $1<d<3$ --
for $d=2$ the field-theoretic predictions\cite{vojta,ssmv03} have been verified
by extensive Quantum Monte Carlo simulations.\cite{troyer,sandvik}

Paranthetically, we note that the fixed point at $\gamma^*$ is unstable
w.r.t. breaking of the underlying SU(2) symmetry: a stable intermediate-coupling
fixed point still exists in the XY case, whereas flow to strong coupling occurs
in the Ising situation.\cite{Zhu,Zarand2}


\section{A magnetic impurity in a $Z_2$ spin liquid}
\label{RG}

As above, we intend to couple a magnetic impurity to the local
order-parameter field of the bulk, i.e.,
by a term
$j_0 \int d\tau \vec{S}(\tau) \cdot \vec\phi(\vec{x}\!=\!0,\tau)$,
but now the low-energy dynamics of $\phi(\vec{x},\tau)$
will be represented in terms of spinon fields $z$ introduced
in Sec.~\ref{noncollinearorder}.

\subsection{Low-energy theory}

It is easy to see that an appropriate choice of a linear combinations of the
spinons defined in Eq.~\eqref{couplingform}, namely
\begin{eqnarray}
z_\uparrow'&=& \frac{1}{\sqrt{2}} \left (z_\uparrow+ i z_\downarrow \right),
\nonumber \\
z_\downarrow'&=&\frac{1}{\sqrt{2}} \left (-i z_\uparrow^*+z_\downarrow \right) \;
\end{eqnarray}
enables us to write expression~\eqref{couplingform} in an alternative, but
very convenient way:
\begin{eqnarray}
\phi(\vec{x}=0)=\mathfrak{Re} \left(\vec{n}_1+i\vec{n}_2 \right)
=z'^{*}_\alpha \frac{\vec{\sigma}_{\alpha\beta}}{2} z'_\beta \; .
\end{eqnarray}
Note that this rotation leaves the bulk action invariant.

The appropriate model to describe a magnetic impurity in a topological spin
liquid with a Z$_2$ gauge structure is thus
\begin{eqnarray}
\mathcal{S}&=& \frac{1}{\gz} \int d^dx  d \tau
\left ( |\partial_x z_\sigma|^2+|\partial_\tau z_\sigma|^2 \right ) \nonumber \\
&+& \int d^dx d\tau \, \kappa(x,\tau) \Big(\sum_\sigma |z_\sigma|^2-1 \Big) \nonumber
\\ &+& j_0 \int d \tau \vec{S} \cdot
z^*_{\sigma} \frac{\vec{\sigma}_{\sigma \sigma'}}{2}z^{\phantom{*}}_{\sigma'}(\vec x = 0) \nonumber
\\&+& \mathcal{S}_{\textrm{Berry}}[\vec{S}]\; ,
\label{equ:bosonickondomodel}
\end{eqnarray}
and was proposed recently by us.\cite{Florens}

The most important feature of model \eqref{equ:bosonickondomodel} is that the impurity
is coupled to a bilinear of the elementary bulk excitations (i.e., the spinons).
This has to be contrasted to the impurity theory with collinear bulk ordering,
where the impurity spin couples {\em linearly} to the magnetic (triplon) modes.
The consequences of this difference will be examined below and turn out to be crucial
for the fate of the impurity near the bulk critical point.
Thus, apart from the strong constraint on the bulk spinons,
the model \eqref{equ:bosonickondomodel} strongly resembles the {\em fermionic} Kondo model
and constitutes its straightforward extension to a bosonic environment.

In the following, we concentrate on the impurity behavior at the bulk critical point,
i.e., for gapless spinons.
In the gapped case, weak impurity coupling will be irrelevant,
and the impurity will decouple from the bulk in the low-energy limit.
However, in analogy to fermionic Kondo models with a gapped density of states
and broken particle--hole symmetry,\cite{Matthiasreview}
a first-order quantum transition will take place upon increasing $j_0$,
beyond which a strong-coupling state is realized.

\subsection{Weak-coupling RG analysis}
\label{sec:weak}

A RG treatment of the action presented in Eq.~\eqref{equ:bosonickondomodel} can
be performed perturbatively in $j_0$.
In addition, the constraint
$\sum_\sigma z^*_\sigma z^{\phantom{\dagger}}_\sigma=1$ is relaxed and treated
on the mean-field level (this is equivalent to giving the spinons a mass shift).
The influence of such bulk interaction terms will be considered in
Sec.~\ref{bulkint}.
At zero temperature, the renormalized spinon mass vanishes at criticality,
and the spinons become soft.
This is the regime we focus on, and which is described by the simpler action:
\begin{eqnarray}
\mathcal{S}&=&\frac{1}{\gz} \int d^dx d\tau
\left (|\partial_x z_\sigma|^2+|\partial_\tau z_\sigma|^2 \right) \nonumber \\
&+& j_0 \int d\tau \vec{S} \cdot
z^*_{\sigma} \frac{\vec{\sigma}_{\sigma \sigma'}}{2}z^{\phantom{*}}_{\sigma'} (\vec x=0)
\nonumber\\
&+& \mathcal{S}_{\textrm{Berry}}[\vec{S}].
\end{eqnarray}
In order to use diagrammatic methods it is convenient to represent the
impurity spin in terms of Schwinger bosons (one could equally well choose to
work with Abrikosov fermions).
Restricting ourselves to an impurity spin $1/2$,
we introduce thus bosonic operators $b_\sigma$ which obey the following constraint
\begin{eqnarray}
Q=\sum_{\sigma} b^{\dagger}_\sigma b^{\phantom{\dagger}}_\sigma=1 \;.
\label{Schwingerconstraint}
\end{eqnarray}
This allows for a faithful impurity spin representation according to
\begin{eqnarray}
\vec{S}=b^{\dagger}_\sigma \frac{\vec{\sigma}_{\sigma \sigma'}}{2}b^{\phantom{\dagger}}_{\sigma'} \; .
\end{eqnarray}
In terms of these bosonic operators, we can formulate the full
impurity action: \cite{barvsstar}
\begin{eqnarray}
\mathcal{S}&=& \frac{1}{\gz}\int d^d x d \tau \, z^*_\sigma
\left( -\partial_x^2-\partial_\tau^2\right)z_\sigma
+\int d\tau \, \overline{b}_\sigma \left (\partial_\tau + \lambda \right)
b_\sigma \nonumber \\
&+& j_0 \int d\tau \, \overline{b}_\sigma
\frac{\vec{\sigma}_{\sigma \sigma'}}{2}b_{\sigma'}z^*_\alpha
\frac{\vec{\sigma}_{\alpha \beta}}{2}z^{\phantom{*}}_\beta (\vec x=0) \nonumber\\
&+& \frac{v_0}{4} \int d\tau \, z^*_\sigma z_\sigma ({\vec x}=0)\; ,
\label{fullmodel}
\end{eqnarray}
where we implement the constraint \eqref{Schwingerconstraint} exactly by taking the
limit $\lambda\to\infty$.\cite{lambda,costi}

In Eq.~\eqref{fullmodel}, we have included a potential scattering term $v_0$ on the impurity
location $\vec x=0$.
In contrast to the fermionic case, such a term is generated perturbatively
in the RG flow. This implies that for finite $j$, $v=0$ cannot be a fixed point
of the RG -- this is intimately related to the bosonic nature of the bulk,
as bosons cannot show particle--hole symmetry.

The RG treatment starts by determining the tree-level scaling dimensions of
the couplings, yielding:
\begin{eqnarray}
\textrm{dim} [j_0] = \textrm{dim} [v_0] = d-2 \equiv
\epsilon \; .
\end{eqnarray}
Both couplings are marginal in two dimensions;
this allows for a controlled RG procedure,
which we shall perform here using the standard field-theoretic scheme.
Renormalized couplings are introduced as
$j_0 =\frac{2}{\gz S_d} \mu^{-\epsilon} \frac{Z_j}{Z_b}j$
and
$v_0 =\frac{2}{\gz S_d} \mu^{-\epsilon} \frac{Z_v}{Z_b}v$,
where $S_d$ denotes $S_d=\frac{2}{\Gamma\left(\frac{d}{2} \right) (4\pi)^{d/2}}$.
At two-loop order (see App.~\ref{appen:bosonrg})
we find the following beta functions
\begin{eqnarray}
\beta (j)
&=&\epsilon j+vj+\frac{j^3}{2}+\mathcal{O}(j^5)\,, \nonumber \\
\beta (v)
&=&\epsilon v+\frac{v^2}{2}+\frac{3}{2}j^2+\mathcal{O}(j^4),
\label{RGeq}
\end{eqnarray}
where $\epsilon=d-2$ for non-interacting bulk bosons.
(This corrects a sign error in the result published in Ref.~\onlinecite{Florens}.)
It is important to note that $\beta(j)$ [$\beta(v)$] does not contain even (odd)
powers of $j$, respectively.
This is the consequence of the invariance of the action \eqref{fullmodel}
w.r.t. the transformation
\begin{eqnarray}
j_0 \to -j_0 \quad \textrm{and} \quad z^{\phantom{\dagger}}_\sigma \to
\epsilon^{\phantom{\dagger}}_{\sigma \sigma'} z^{*}_{\sigma'}
\label{equ:trafo}
\end{eqnarray}
which also implies that ferromagnetic and antiferromagnetic impurity
coupling are equivalent.
(Note again here that the $z$ are not canonical bosons -- the corresponding
operators simply commute.)

\begin{figure}[t]
\begin{center}
\includegraphics[width=0.48\textwidth]{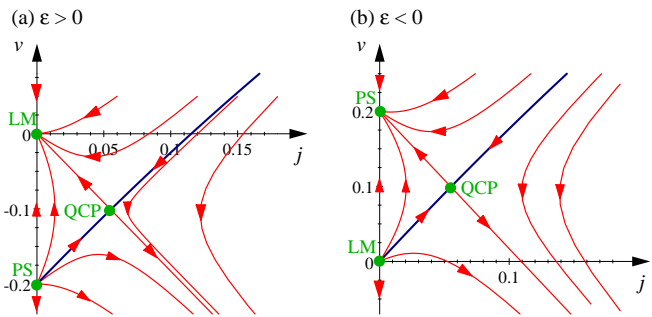}
\end{center}
\caption{
RG flow of the spinon Kondo model \eqref{fullmodel}
at the bulk critical point.
(a) $\epsilon>0$; (b) $\epsilon<0$.
($\epsilon = d-2$ for non-interacting spinons;
note that $\epsilon$ may be non-zero in $d=2$ owing to bulk spinon interactions.)
QCP denotes the boundary quantum critical point and the blue line is the separatrix,
separating a strongly coupled phase from either a local moment (LM) regime or a
potential scattering (PS) a zero Kondo coupling.
The curves have been calculated from \eqref{RGeq} for $\epsilon=\pm 0.1$.
Note that the flow is symmetric w.r.t. $j \leftrightarrow -j$.
}
\label{fig:flow_bose}
\end{figure}

The solution of the two-loop flow equations is depicted in Fig.~\ref{fig:flow_bose}.
The weak-coupling RG flow has four fixed points:
the trivial (Gaussian) fixed point LM with $(v^*,j^*)=(0,0)$;
a potential scattering fixed point PS, given by $(v^*,j^*)=(-2\epsilon,0)$;
and a pair of critical fixed point, located at
$(v^*,j^*)=(-\epsilon -\frac{\epsilon^2}{6}, \pm\frac{\epsilon}{\sqrt{3}})$
and denoted QCP.

Let us quickly discuss the physical implications of the present RG flow
(but note that certain changes will arise from the consideration of
bulk interactions, see below).

For $\epsilon>0$ and $j\neq 0$, the flow indicates two stable phases, namely
the local-moment phase (LM), and a strong-coupling phase at large $|j|$
(the Hamiltonian is symmetric under $j_0 \to -j_0$) and large negative $v$.
The two phases are separated by a continuous quantum phase transition,
controlled by QCP, with a correlation length exponent given by
\begin{eqnarray}
\frac{1}{\nu}=|\epsilon|-\frac{\epsilon^2}{12}+\mathcal{O}(\epsilon^3),
\label{invnu1}
\end{eqnarray}
characterizing the vanishing crossover energy scale near criticality.
The pure potential scattering problem, $j=0$, has two phases as well,
which can be easily understood from the exact local spinon Green function:
\begin{eqnarray}
\label{deno}
G(i\nu) & = & \frac{G_0(i\nu)}{1+(v_0/4)G_0(i\nu)}\\
\label{bare}
\mr{where} \;\; G_0(i\nu) & = & \int \frac{d^dk}{(2\pi)^d} \frac{g}{\nu^2+k^2}\,.
\end{eqnarray}
Values $v>-2\epsilon$ flow to zero (the denominator in Eq.~(\ref{deno}) is regular),
whereas $v<-2\epsilon$ induces a bound state (a pole appears in the Green
function of the bosons).
The physical properties of LM are that of a decoupled impurity,
i.e., susceptibility and entropy obey
$\chi_{\rm imp} = 1/(4T)$, $S_{\rm imp} = \ln 2$
in the low-temperature limit.
The strong-coupling regime cannot be analyzed using the present RG;
a combination of large-$N$ methods and strong-coupling models indicates
that true screening can occur (for both signs of $j_0$),
with a finite impurity susceptibility and vanishing entropy.\cite{Florens}
Finally, at the critical point (QCP) properties similar to
Eq.~\eqref{fracspin} obtain.

For $\epsilon<0$, the weak-coupling regime is unstable, and the flow
generically is towards strong coupling -- this can be related
to the bare boson density of states, Eq.~(\ref{bare}), diverging as $\nu\to 0$.
Nevertheless, there is still a QCP, which separates the strong-coupling
(screened) regime from a pure potential-scattering phase
at positive $v$ (controlled by a now stable PS fixed point).

For $\epsilon=0$ no quantum phase transition occurs, and
there is logarithmically slow flow to strong coupling
(see below for corrections to this picture coming from bulk interactions).

Overall, the RG flow is strikingly different from an impurity which is coupled to
gapless spin-$1$ bosons (Sec.~\ref{vojtabura}),
where the coupling $\gamma$ flows towards a {\em stable}
intermediate-coupling fixed point.

\subsection{The role of bulk interactions}
\label{bulkint}

Within the above weak-coupling RG the bulk interactions
(represented by the constraint $\sum_\sigma |z_\sigma|^2=1$)
were only treated at the mean-field level.
Unfortunately, a consistent RG fully including the bulk interactions cannot be performed:
the appropriate bulk non-linear sigma model is perturbatively accessible near $d=1$
(while softening the bulk interaction allows an expansion in $3-d$, see below),
but expansions around both $d=1$ and $d=3$ are incompatible with the bare scaling
dimension of the impurity coupling $j_0$, which is marginal in $d=2$.
(This is in contrast to the situation in Sec.~\ref{vojtabura} where
{\em both} bulk and boundary couplings were marginal at $d=3$.)

The treatment of bulk interactions in the present spinon case is, however,
physically important, both at weak coupling (to be discussed below)
and at strong coupling -- this was discussed in some detail in
Ref.~\onlinecite{Florens}, but requires further work which is beyond the
scope of this paper.

Let us qualitatively discuss the role of bulk spinon interactions at weak
coupling. The scaling dimensions of $j$ and $v$ are determined by the local bulk
spin correlations. In the interacting bulk theory, this decay is given as
$\chi (\tau) \propto 1/\tau^{d-1+\eta}$ where $\eta$ is the anomalous dimension.
(Note that $\eta=1$ for non-interacting spinons, whereas $\eta=0$ in the
case of a standard $\phi^4$ theory controlled by a Gaussian fixed point.)
This implies that
$\textrm{dim}[j_0] = \textrm{dim}[v_0] = (d-3+\eta)/2$, i.e.,
the $\epsilon=d-2$ in the RG equations~(\ref{RGeq}) should be replaced by
$\epsilon=(d-3+\eta)/2$.
The QCP in the incommensurate magnet has $\eta>1$ in $d=2$ (for details
see Refs.~\onlinecite{isakov,chubukov}), i.e., a boundary quantum phase transition as shown
in Fig.~\ref{fig:flow_bose}(a) is indeed possible for the physical
two-dimensional case.

We note that this analysis (i.e. including $\eta$ in the scaling dimensions of
$j$ and $v$) is in principle correct at one-loop order;
at higher orders interference processes between bulk and boundary interactions
can be expected.
No systematic treatment of those is possible in the present formulation.
Therefore, we propose a more rigorous way to incorporate bulk
interactions, which leads us to a bosonic version of the Anderson
impurity model.

\subsubsection{Derivation of the bosonic Anderson model}
\label{sec:boseanderson}

Our goal is to find a representation of the model \eqref{fullmodel}
which allows to perturbatively control both bulk and boundary interactions
at the same time.
To this end, we reformulate the impurity part, i.e., switch from a Kondo
to an Anderson impurity model -- such a scheme has proven extremely successful
for the pseudogap Kondo model.\cite{Fritz}

Two transformations will convert the original model \eqref{fullmodel}
into a new one, where bulk and boundary interactions have
a common critical dimension:
(a) an inverse Schrieffer-Wolff transformation (in analogy to the pseudogap Kondo case),
which transforms the Kondo into an Anderson model and renders the hybridization
between impurity and bulk marginal in $d=3$;
(b) replacing the hard constraint on the spinons by a ``soft'' self-interaction
of the type $u_0(|z_\sigma|^2-1)^2$ -- this term is marginal in $d=3$
as well.
(As will become clear, step (a) requires the existence of a strong-coupling
singlet state with screening in the original model.)
With steps (a) and (b) we are led to propose the following field theory:
\begin{eqnarray}
\mathcal{S} &=& \int d^dx d \tau
\left[ \frac{1}{\gz} \left (|\partial_x z_\sigma|^2+|\partial_\tau z_\sigma|^2 \right)
+ u_0 z^{*}_\sigma z^{\phantom{*}}_\sigma z^{*}_{\sigma'}
z^{\phantom{*}}_{\sigma'} \right]
\nonumber \\
& & + \int d\tau \left[ \overline{b}_0 \left (\partial_\tau+\lambda
+ \epsilon_0 \right)b_0 +\overline{b}_\sigma
\left (\partial_\tau +\lambda \right) b_\sigma \right]
\nonumber \\
&+& w_0 \int d\tau \left[ \overline{b}_\sigma
b_0 z^{\phantom{*}}_\sigma+\textrm{h.c.}\right].
\label{bosoAnd}
\end{eqnarray}
We have left out a bulk mass term $\propto z_\sigma^2$, as we are interested in the
bulk critical point.
The impurity is now a three-level system
(similar to an Anderson model with infinite repulsion),
represented by three auxiliary particles
$b_\sigma$ ($\sigma=\uparrow,\downarrow$) and $b_0$ -- note
that $b_0$ may be interpreted as the empty impurity state.
The impurity transition is tuned by the value of $\epsilon_0$:
for $\epsilon_0 > 0$ the impurity is in a doublet state (i.e. unscreened)
whereas for $\epsilon_0 < 0$ the state is a singlet
(these statements apply for $w_0\to 0$, finite $w_0$ will shift the impurity critical point).
The Hilbert space dimension is enforced by the constraint
$Q=\sum_\sigma b^{\dagger}_\sigma b^{\phantom{\dagger}}_\sigma + b^{\dagger}_0
b^{\phantom{\dagger}}_0=1$.
Technically, this is again implemented via a chemical potential $\lambda \to
\infty$.\cite{lambda,costi}

The action \eqref{bosoAnd} incorporates three main ingredients:
(i) it contains an interaction term $u_0$ in the bulk, which is marginal in $d=3$;
(ii) the hybridization $w_0$ is marginal in $d=3$ as well;
(iii) in the ``Kondo'' limit, i.e., for a large positive singlet energy $\epsilon_0$,
a Schrieffer-Wolff transformation allows to integrate out the $b_0$ field,
and leads to the previous bosonic Kondo interaction~(\ref{fullmodel}).
The model~(\ref{bosoAnd}) at $d=3$ is thus strongly reminiscent of the pseudogap Anderson
impurity model (with infinite impurity Coulomb energy) and linear density of
states.\cite{Fritz1}
(Another bosonic Anderson model was recently introduced by Lee and Bulla,\cite{lee}
which, however, displays different physics, as the model of Ref.~\onlinecite{lee}
is spinless, but with finite on-site repulsion.)

We will proceed with deriving the RG flow for the Anderson model \eqref{bosoAnd}.

\subsubsection{RG analysis of the bosonic Anderson model}
\label{sec:ande}

The RG flow for both the bulk and impurity interactions, $u_0$ and $w_0$,
can be perturbatively controlled in $\epsilon=\frac{3-d}{2}$.
This allows a consistent calculation of the critical properties of the
fully interacting bosonic impurity model.
The details of this calculation are presented in App.~\ref{appen:bosonicandersonrg}.

We introduce renormalized couplings according to
$w_0=\mu^{\epsilon}\sqrt{\frac{2 }{\gz S_d}} \frac{Z_w}{\sqrt{Z_{b_0}Z_{b_\sigma}Z_z}}w$ and
$u_0=\mu^{2\epsilon}\frac{4}{S_d \gz^2}\frac{Z_u}{Z_z^2} u$.
Placing ourselves at both the bulk and boundary critical points (!),
we find the following flow equations to one-loop order:
\begin{eqnarray}
\beta(w) &=& \epsilon w-\frac{3}{2}w^3, \nonumber \\
\beta(u) &=& 2 \epsilon u-12 u^2.
\end{eqnarray}
These RG equations describe the non-trivial flow of both bulk and
boundary interactions at the critical point in $d<3$.
Higher-order contributions (not shown here) contain the above-mentioned interference
processes and induce a feedback of the bulk interaction $u$ into the impurity flow.

We refrain from a more detailed analysis at this point:
qualitatively, the RG approaches in Secs.~\ref{sec:weak} and \ref{sec:ande}
yield similar results; a quantitative comparison is difficult as the
two expansions are around a lower critical and upper critical dimension of the
impurity problem, respectively.
Let us, however, emphasize that we have demonstrated how a systematic expansion,
analogous to the one in Ref.~\onlinecite{vojta} for impurity models
in collinear magnets (i.e. treating bulk and boundary interactions on equal footing)
can be set up in the non-collinear case as well.


\section{A magnetic impurity in a two-component canonical Bose gas}
\label{canonical}

In this last part, we briefly introduce and analyze the problem of a two-state
system\cite{Zoller} (described by a magnetic pseudospin impurity) coupled to a
two-component non-relativistic Bose gas.\cite{williams}
This extends the kind of bosonic impurity models,
introduced above in the context of quantum antiferromagnets,
to the field of cold atomic gases.

\subsection{The model}

Technically, the main difference to the situation in the previous section
is that the bulk now consists of {\em canonical} bosons.
The action assumes the form:
\begin{eqnarray}
\mathcal{S}&=& \int d^dx d \tau \,
\overline{a}_\sigma (x,\tau) \left (\partial_\tau-\partial_x^2
+m \right)a_\sigma \nonumber \\
&+& j_0 \int d \tau \vec{S} \cdot
\overline{a}_\sigma \frac{\vec{\sigma}_{\sigma \sigma'}}{2}a_{\sigma'}({\vec x}=0) \nonumber\\
&+& \frac{v_0}{4} \int d\tau \, \bar{a}_\sigma a_\sigma ({\vec x}=0) \nonumber\\
&+& \mathcal{S}_\textrm{Berry}[\vec S] \; ,
\label{canmodel}
\end{eqnarray}
where we consider canonical bosonic particles, $a_\sigma$, with two internal degrees of
freedom ($\sigma=\uparrow,\downarrow$ here) and a mass $m$ (i.e., chemical potential).
In the field of ultracold trapped gases, usually performed with fixed particle
number, $m\to 0$ in the low-temperature limit.
Interactions among bulk bosons are neglected; to lowest order, they lead to a
temperature-dependent renormalization of $m$.
The impurity couples to the bosonic spin
density at site ${\vec x}=0$ with a Kondo-type exchange interaction, but an
Anderson-like Hamiltonian similar to Eq.~(\ref{bosoAnd}) could be considered as well.

\subsection{Weak-coupling analysis}

In contrast to spinon-based models, we realize that there is no invariance
under the transformation $j_0 \rightarrow -j_0$, i.e.,
ferromagnetic and antiferromagnetic impurity coupling are no longer equivalent.
However, as in the spinon case,
a potential scattering term is generated during the flow even for $v_0=0$.

In the remainder, we focus on the weak-coupling analysis in the massless
case, $m=0$.
(In the massive case, $m>0$ at $T=0$, we again expect a first-order
transition as function of $j_0$.)
A tree-level dimensional analysis yields
\begin{eqnarray}
\textrm{dim}[j_0]=\textrm{dim}[v_0]=d-2\equiv \epsilon ,
\end{eqnarray}
similar to the model in Sec.~\ref{sec:weak}.
We thus introduce dimensionless couplings according to
$j_0 =\mu^{-\epsilon} \frac{1}{S_d}\frac{Z_v}{Z_b}j$ and
$v_0 =\mu^{-\epsilon} \frac{1}{S_d}\frac{Z_v}{Z_b}v$
and find to one-loop order
\begin{eqnarray}
\beta(j) &=& \epsilon j-\frac{j^2}{2}+\frac{v j}{2},\nonumber \\
\beta(v) &=& \epsilon v+\frac{v^2}{4}+\frac{3}{4}j^2 \; .
\label{canRGeq}
\end{eqnarray}
Remarkably, this result is exact to {\em all} orders in perturbation theory (this was
previously shown for non-relativistic $\phi^4$-theory with local
interactions\cite{Sachdevbook,bergmann}). Technically, the reason for that lies in the
diagrammatic structure of the problem, where only ladder-like diagrams can be different
from zero in the limit $T=0$. Higher-order terms are cancelled by the appropriate
counterterms. Physically, the exactness of Eq.\eqref{canRGeq} is tied to the fact that
the expansion is performed around a trivial vacuum state.

\begin{figure}[t]
\begin{center}
\includegraphics[width=0.47\textwidth]{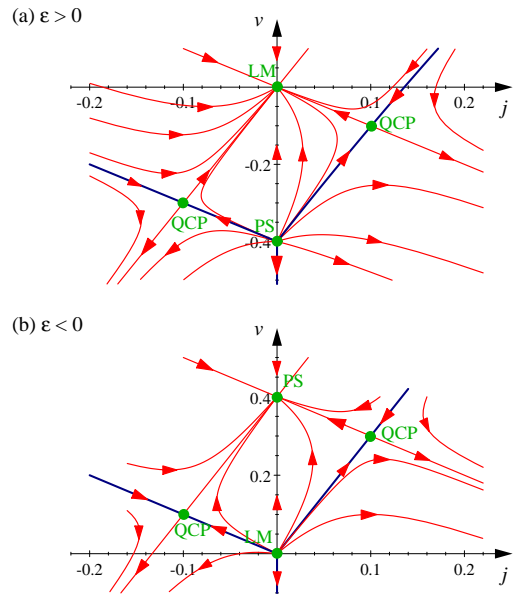}
\end{center}
\caption{
RG flow of the Kondo model with massless canonical bosons,  Eq.~\eqref{canmodel}.
(a) $\epsilon>0$; (b) $\epsilon<0$.
($\epsilon = d-2$ for non-interacting bosons;
again, $\epsilon$ may be non-zero in $d=2$ due to bulk interactions.)
The curves have been calculated from \eqref{canRGeq} for $\epsilon=\pm 0.1$.
}
\label{fig:flow_canonicalbose}
\end{figure}

The RG flow, derived from Eq.~\eqref{canRGeq}, is depicted in
Fig.~\ref{fig:flow_canonicalbose}.
For positive (antiferromagnetic) $j$, it is qualitatively similar to
the one of the spinon model in Sec.~\ref{sec:weak}.
From Eq.~\eqref{canRGeq} one finds four fixed points:
the local-moment (LM) fixed point at $(v^*,j^*)=(0,0)$,
a potential-scattering (PS) fixed point at $(v^*,j^*)=(-4\epsilon,0)$,
and two critical fixed points at
$(v^*,j^*)=(-\epsilon,\epsilon)$ and
$(v^*,j^*)=(-3\epsilon,-\epsilon)$.
For both QCP, the correlation length exponent is
\begin{eqnarray}
\frac{1}{\nu}=|\epsilon| \; ,
\end{eqnarray}
which again is exact to all orders in perturbation theory.

The structure of the phase diagram for both positive and negative $\epsilon$ is
thus similar to the spinon case in Sec.~\ref{sec:weak}.
However, ferromagnetic and antiferromagnetic $j_0$ are no longer equivalent,
suggesting the existence of two distinct strong-coupling phases for
$j<0$ and $j>0$ (where true screening is likely to occur only in the latter case).

For $\epsilon=0$ and $j>0$, logarithmic flow to strong coupling occurs,
with a screening temperature given by $\ln T^\ast \propto -1/j_0$
as in the conventional Kondo problem.

Thus, the situation of a two-level impurity coupled to a two-component Bose gas,
with a pseudospin--pseudospin coupling, allows for three distinct phases (given $j_0\neq 0$):
a weakly coupled (unscreened) phase and two strong-coupling phases.
The present analysis can be easily generalized to an arbitrary number
of pseudospin components or larger effective spin sizes;
at weak coupling no qualitative changes are expected.
However, the precise nature of the ground state at strong coupling depends on
microscopic parameters of the Bose gas, such as spin size and the value of the
repulsion between bosonic atoms.\cite{Florens}
More elaborate strong-coupling methods are required to address this issue.

Last not least, let us point out another remarkable feature of the above model
\eqref{canmodel}:
its RG flow is equivalent to a fermionic Kondo problem with full
band asymmetry, i.e, with a chemical potential tuned to the band edge.
This was considered previously,\cite{Zawadowski} and we comment on this point
in App.~\ref{appen:fermions}.


\section{Conclusion}

In this paper, we have introduced a new class of quantum impurity models,
describing the dynamics of a two-level system (spin) coupled to an external bosonic
spin-carrying environment.
Two different situations have been studied in the weak-coupling regime,
namely a spin $1/2$ magnetic impurity in a class of $Z_2$ spin liquids,
and a local defect in a two-component Bose gas.

For massless bulk particles, we have derived the renormalization group flow
of the impurity coupling constants.
While some differences in the RG flow appear between the above two realizations,
the following generic features appear:
(I) an impurity quantum phase transition is possible which separates a
local-moment regime (decoupled impurity) from a strongly coupled impurity;
(II) in contrast to a magnetic impurity coupled to a collinear bulk magnet
described by a standard $\phi^4$ theory, there is no stable intermediate-coupling
fixed point, but instead run-away flow to strong coupling.
(For gapped bulk particles, a first-order transition will generically appear,
in analogy to gapped fermionic Kondo models.)
At strong coupling, the nature of the ground state depends on
microscopic details (and was not investigated further here), but quenching
of the impurity entropy may obtain.\cite{Florens}

In summary, our results underline that the dynamics of impurity degrees of freedom
depends crucially on the nature of the elementary excitations of the surrounding
environment.
Hence, impurity properties may be used as probes for exotic bulk excitations.
On the experimental side, Cs$_2$CuCl$_4$ has been proposed to be a realization
of a spin-1/2 frustrated antiferromagnet on an anisotropic triangular lattice,
which possibly features spinon excitations.\cite{coldea0,coldea}
Studying the physics of dilute magnetic impurities in this material,
e.g. the temperature dependence of NMR Knight shifts or of the impurity contribution
to the susceptibility, will clearly help to resolve the nature of the
bulk magnetic state.


\acknowledgments

We thank S. Sachdev, T. Senthil and G. Zar\'and for valuable discussions, and K. Damle
for a collaboration at the early stage of this work. Furthermore, we would like to
acknowledge the creators of Jaxodraw\cite{Jaxodraw}, which simplified the drawing of the
Feynman diagrams significantly.
This research was supported by the DFG through the Center
for Functional Nano\-structures (Karlsruhe) and SFB 608 (K\"{o}ln), as well as through
the Virtual Quantum Phase Transitions Institute in Karlsruhe.


\appendix

\section{Weak-coupling RG for the bosonic Kondo model}
\label{appen:bosonrg}

In this appendix we derive the RG equations for the bosonic Kondo model with bulk spinons
(the derivation is kept in a form which is also appropriate to treat the case of
canonical bosons as well as electrons; the only occurring modifications manifest
themselves in the values of the corresponding Feynman diagrams). The RG equations of
bosons are very different form those of electrons, which is the reason why we present a
rather detailed analysis. The crucial difference resides in the fact that during the
renormalization process a potential scattering term is generated. In order to properly
deal with this complication we start with a generalized version of the model proposed in
Eq.~\eqref{equ:bosonickondomodel} which allows for fully spin-anisotropic couplings. This
enables us to deal with the Kondo interaction and the potential scattering term at a time
without worrying about newly generated terms.

The action we analyze has the following very general form
\begin{eqnarray}
\mathcal{S}&=& \frac{1}{\gz} \int d^dx d \tau
\sum_{\sigma} z^*_{\sigma} \left( -\partial_\tau^2-\partial_x^2\right)z_\sigma  \nonumber \\  &+& \sum_\sigma \int d \tau \overline{b}_\sigma \left (\partial_\tau + \lambda \right)b_\sigma \nonumber \\
&+& \alpha_0 \int d \tau \sum_{\sigma \neq \sigma'} z^*_{\sigma} z_{\sigma'}
\overline{b}_{\sigma'} b_\sigma \nonumber \\
& +& \beta_0 \int d \tau \sum_{\sigma} z^*_{\sigma} z_{\sigma}
\overline{b}_{\sigma} b_\sigma \nonumber \\
&+&\gamma_0 \int d \tau \sum_{\sigma \neq \sigma'} z^*_{\sigma} z_{\sigma}
\overline{b}_{\sigma'} b_{\sigma'}  \; .
\end{eqnarray}

\subsubsection{Weak-coupling RG procedure}

\begin{figure}
\begin{center}
\includegraphics[width=0.48\textwidth]{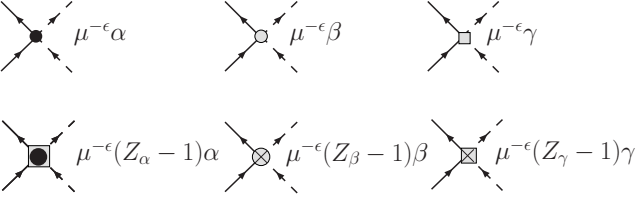}
\end{center}
\caption{
Graphical representation of all the occurring vertices; $\alpha$ is denoted as
a black circle, its counterterm is represented by a big black circle in a box; the small
grey dot and the big grey dot with a cross represent $\beta$ and its counterterm; the
small (big) grey box (grey box with a cross) represents $\gamma$ (the counterterm of
$\gamma$).
}
\label{vertices}
\end{figure}

The tree-level scaling dimensions of the general coupling constants are
\begin{eqnarray}
\textrm{dim}[\alpha_0]=\textrm{dim}[\beta_0]=\textrm{dim}[\gamma_0]=d-2=\epsilon \; .
\end{eqnarray}
Based on this analysis we introduce dimensionless couplings according to
\begin{eqnarray}
\alpha_0 &=&\mu^{-\epsilon} \frac{2}{\gz S_d} \frac{Z_\alpha}{Z_b}\alpha \; , \nonumber \\
\beta_0  &=&\mu^{-\epsilon} \frac{2}{\gz S_d} \frac{Z_\beta}{Z_b} \beta \; \nonumber \\
\gamma_0 &=&\mu^{-\epsilon} \frac{2}{\gz S_d} \frac{Z_\gamma}{Z_b} \gamma \; ,
\end{eqnarray}
where $\mu$ denotes an arbitrary but fixed energy scale (the renormalization scale) and $S_d=\frac{\Omega_d}{(2\pi)^d}$, where $\Omega_d=\frac{2\pi^{d/2}}{\Gamma(d/2)}$ with $\Gamma(x)$ being the Euler function. We
furthermore state that during the course of renormalization the Schwinger boson field
parametrizing the spin is renormalized according to
\begin{eqnarray}
b_R=Z_b^{-1}b \;.
\end{eqnarray}
We do not have to introduce renormalized fields for the bulk degrees of freedom since the
influence of  a single impurity on the bulk is an effect of the order ${\cal O}(1/N)$, which does
not renormalize bulk degrees of freedom. The prefactor $\frac{2}{\gz S_d}$ can trivially be
taken care of by rescaling the fields in Eq.~\eqref{equ:bosonickondomodel} according to
\begin{eqnarray}
z_{\sigma} = \sqrt{\frac{ \gz S_d}{2}} z'_\sigma
\end{eqnarray}
which has the only advantage of getting rid of bothersome factors in the course of the
calculation. The generalized renormalized action (with appropriate counterterms) reads:
\begin{eqnarray}
\mathcal{S}_R&=&  \frac{S_d}{2 } \int d^dx d \tau \sum_{\sigma} z^*_{\sigma}
\left( -\partial_\tau^2-\partial_x^2\right)z_\sigma\nonumber \\
&+&\mu^{-\epsilon} \alpha \int d \tau \sum_{\sigma \neq \sigma'}
z^*_{0,\sigma} z_{0,\sigma'} \overline{b}_{\sigma'} b_\sigma \nonumber \\
& +& \mu^{-\epsilon} \beta \int d \tau \sum_{\sigma} z^*_{0,\sigma} z_{0,\sigma}
\overline{b}_{\sigma} b_\sigma \nonumber \\
&+& \mu^{-\epsilon} \gamma \int d \tau \sum_{\sigma \neq \sigma'}
z^*_{0,\sigma} z_{0,\sigma} \overline{b}_{\sigma'} b_{\sigma'} \nonumber \\
&+& \int d \tau \sum_\sigma \overline{b}_\sigma
\left ( \partial_\tau+\lambda \right) b_\sigma
+\sum_\sigma \int d \tau (Z_b-1) \overline{b}_\sigma \partial_\tau b_\sigma \nonumber \\
&+& \mu^{-\epsilon} \alpha (Z_\alpha-1) \int d \tau \sum_{\sigma \neq \sigma'}
z^*_{0,\sigma} z_{0,\sigma'} \overline{b}_{\sigma'} b_\sigma \nonumber \\
& +& \mu^{-\epsilon} \beta (Z_\beta-1) \int d \tau \sum_{\sigma} z^*_{0,\sigma} z_{0,\sigma}
\overline{b}_{\sigma} b_\sigma \nonumber \\
&+& \mu^{-\epsilon} \gamma (Z_\gamma-1)\int d \tau \sum_{\sigma \neq \sigma'}
z^*_{0,\sigma} z_{0,\sigma} \overline{b}_{\sigma'} b_{\sigma'}\; .\nonumber \\
\end{eqnarray}
This is the appropriate starting point to iteratively construct the counterterms by
demanding the action to be finite at a certain arbitrary but fixed renormalization point.
The graphical representation of the vertices and the counterterms is shown in
Fig.~\ref{vertices}. The vertex $\alpha$ is denoted as a small black dot whereas the
associated counterterm $\left( Z_\alpha -1 \right) \alpha$ is represented as a big black
dot in a box. The interaction vertex $\beta$ and its counterterm $\left (Z_\beta-1
\right) \beta$ are represented as a small grey dot and a big grey dot with a cross in the
interior, whereas $\gamma$ and the associated counterterm $\left (Z_\gamma-1 \right)
\gamma$ are graphically represented as small grey box and a big grey box with a cross
inside, respectively.
In the diagrammatic expansion this enforces us to consider 6 interaction vertices for a consistent treatment.

The next step is to specify the renormalization conditions which will give the
theory a finite limit once the cutoff is sent to infinity according to
\begin{eqnarray}
\Gamma_b^2(i \nu-\lambda=i\overline{\nu}=0)&=&0 \; ,  \nonumber \\  \frac{\partial}{\partial i\overline{\nu}}
\Gamma_b^2 (i \overline{\nu}) \bigg|_{i\overline{\nu}=\mu}&=&1 \; , \nonumber \\
\Gamma^4_{ss'ss'}\bigg|_{R}&=&\mu^{-\epsilon}\alpha \; , \nonumber \\ \Gamma^4_{ssss}\bigg|_{R}
&=&\mu^{-\epsilon}\beta \; , \nonumber \\  \Gamma^4_{sss's'} \bigg|_{R}&=&\mu^{-\epsilon}\gamma\; ,  \nonumber \\
\label{equ:rgconditions}
\end{eqnarray}
where the subscript denotes the appropriate renormalization point, which we do not specify here. The first line fixes the impurity propagator, whereas the second line fixes the vertex functions. We once again stress the fact that the bulk properties will not receive singular corrections in any order of perturbation theory.

\subsubsection{Result to one-loop order}
\begin{figure}
\begin{center}
\includegraphics[width=0.33\textwidth]{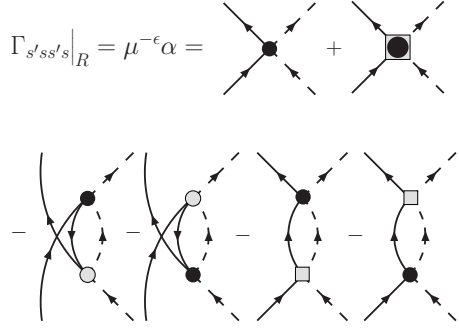}
\end{center}
\caption{
Graphical representation of the renormalization condition shown in
Eq.~\eqref{equ:rgconditions} determining the one-loop counterterm $(Z_\alpha-1)^{(1)}$ (big black
dot in box).
}
\label{fig:gammaalpha1loop}
\end{figure}

\begin{figure}
\begin{center}
\includegraphics[width=0.33\textwidth]{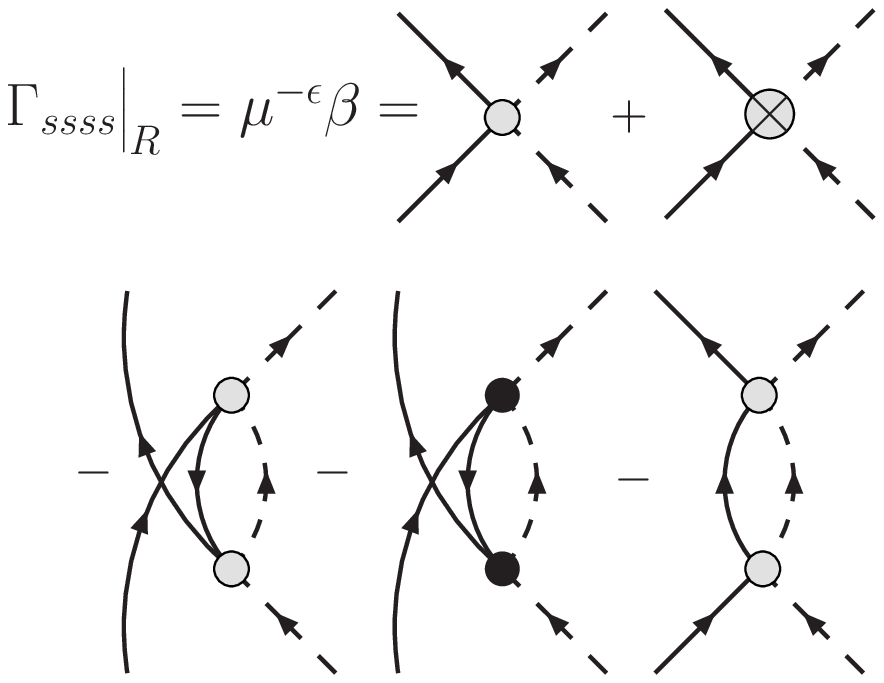}
\end{center}
\caption{
Graphical representation of the renormalization condition shown in
Eq.~\eqref{equ:rgconditions} determining the one-loop counterterm $(Z_\beta-1)^{(1)}$ (big grey
dot with a cross).
}
\label{fig:gammabeta1loop}
\end{figure}

\begin{figure}
\begin{center}
\includegraphics[width=0.33\textwidth]{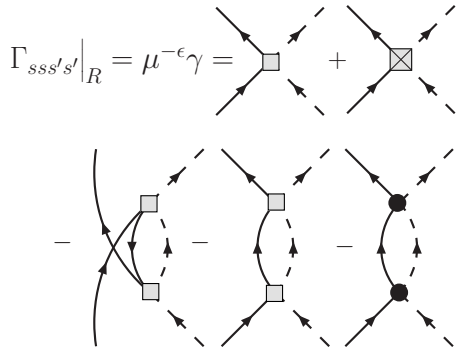}
\end{center}
\caption{
Graphical representation of the renormalization condition shown in
Eq.~\eqref{equ:rgconditions} determining the one-loop counterterm $(Z_\gamma-1)^{(1)}$ (big grey
box with a cross).
}
\label{fig:gammagamma1loop}
\end{figure}

To lowest loop order we can concentrate on the renormalization of the vertex functions since there is no propagator renormalization at one loop level (they will play a role at two loop level).
To lowest loop order there are no propagator renormalizations. We choose a graphical representation of the
renormalization conditions fixed in Eq.~\eqref{equ:rgconditions}. The
renormalization conditions for the vertex functions are shown in
Fig.~\ref{fig:gammaalpha1loop}, Fig.~\ref{fig:gammabeta1loop}, and
Fig.~\ref{fig:gammagamma1loop}. Structurally there are only two types of
diagrams occurring, whose most divergent part in an expansion in $1/\epsilon$ yields
\begin{eqnarray}
\mu^{2 \epsilon} \int_0^{\infty} dx \frac{x^{\epsilon+1}}{x(1+x)}
=-\mu^{2 \epsilon} \left (\frac{1}{\epsilon}+\mathcal{O}(\epsilon^0) \right) \; .
\end{eqnarray}
By dressing the integral values with the appropriate interaction vertices we can
determine the $Z$ factors to lowest order as
\begin{eqnarray}
Z_\alpha^1&=&-\left(2\beta +2\gamma \right) \frac{1}{\epsilon} \; , \nonumber \\
Z_{\beta}^1&=&-\left ( \frac{\alpha^2}{\beta}+2 \beta  \right)  \frac{1}{\epsilon}\; , \nonumber \\
Z_{\gamma}^1&=&-\left ( \frac{\alpha^2}{\gamma}
+2 \gamma \right)  \frac{1}{\epsilon} \; .
\end{eqnarray}

\subsubsection{Result to two-loop order}
\begin{figure*}
\begin{center}
\includegraphics[width=0.9\textwidth]{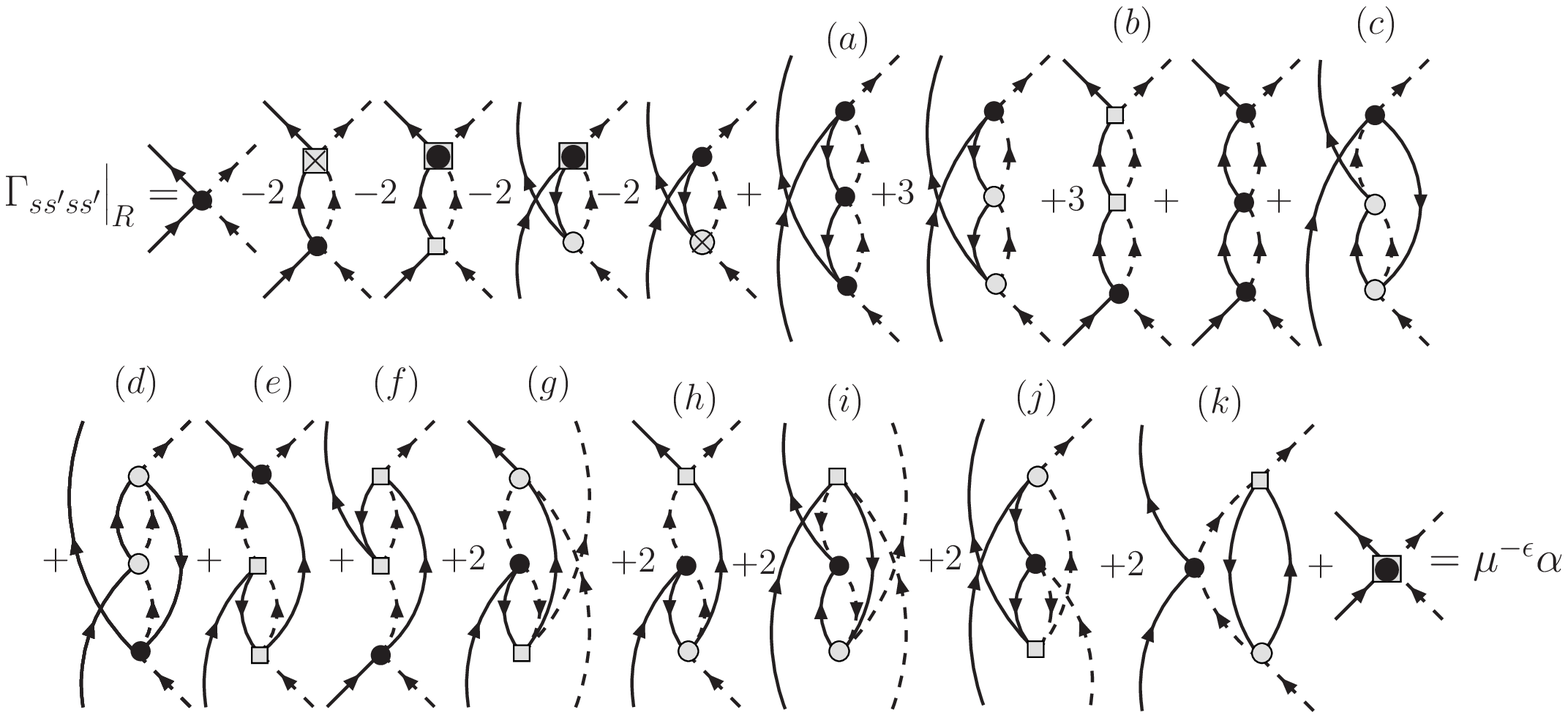}
\end{center}
\caption{
Graphical recursion relation for the two-loop contribution
to the renormalization factor $(Z_\alpha-1)^{(2)}$ (last term).
Note that the counterterms appearing as internal vertices in the diagrams
take their one-loop values $(Z_\alpha-1)^{(1)}\alpha$, $(Z_\beta-1)^{(1)}\beta$, $(Z_\gamma-1)^{(1)}\gamma$.
(a)-(k) denote topologically distinct diagrams.}
\label{fig:Gammaalpha}
\end{figure*}

\begin{figure*}
\begin{center}
\includegraphics[width=0.9\textwidth]{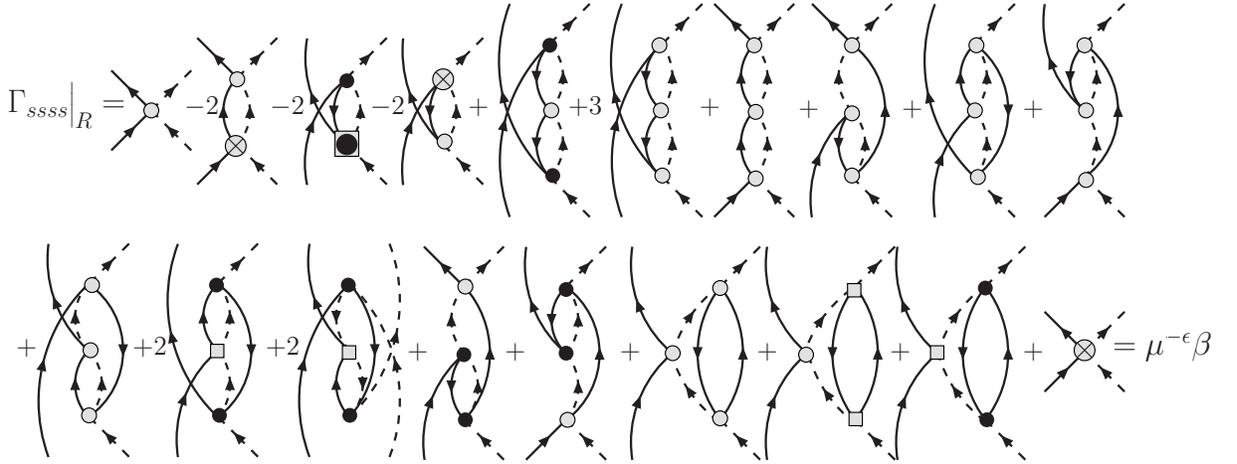}
\end{center}
\caption{
Graphical recursion relation for the two-loop contribution
to the renormalization factor $(Z_\beta-1)^{(2)}$ (last term).
Note that the counterterms appearing as internal vertices in the diagrams
take their one-loop values $(Z_\alpha-1)^{(1)}\alpha$, $(Z_\beta-1)^{(1)}\beta$, $(Z_\gamma-1)^{(1)}\gamma$.
}
\label{fig:Gammabeta}
\end{figure*}

In order to derive the $Z$ factors to two-loop order we only calculate the numerical
values of the topologically distinct diagrams shown in Fig.~\ref{fig:Gammaalpha}. The
actual vertex corrections are obtained by properly dressing the numerical value of the
integrals with the corresponding interaction vertices. All possible diagrams contributing
to the two-loop renormalization factors are shown in Figs.~\ref{fig:Gammaalpha}
and~\ref{fig:Gammabeta}. The diagrams contributing to $Z_\gamma$ are not shown
explicitly, since they can be obtained from the diagrams contributing to $Z_\beta$ by
replacing grey dots by grey boxes. We start with the trivial diagrams $(a)$ and $(b)$
which are just the squares of the one-loop diagrams
\begin{eqnarray}
(a)=(b)= \mu^{2\epsilon}\left( \frac{1}{\epsilon^2} +\mathcal{O}(\epsilon^0) \right) \; .
\end{eqnarray}
The diagrams $(c) ... (j)$ are given by
\begin{eqnarray}
(c...j)= \mu^{2\epsilon} \int_0^{\infty} d x \frac{x^\epsilon (x+2)^\epsilon}{1+x} \int_0^{\infty} d y \frac{y^{\epsilon}}{1+y}\; ,
\end{eqnarray}
which to leading order yields
\begin{eqnarray}
(c....j)= \mu^{2\epsilon}
\left( \frac{1}{2\epsilon^2}+\mathcal{O}(\epsilon^0)\right).
\end{eqnarray}

\begin{figure}
\begin{center}
\includegraphics[width=0.48\textwidth]{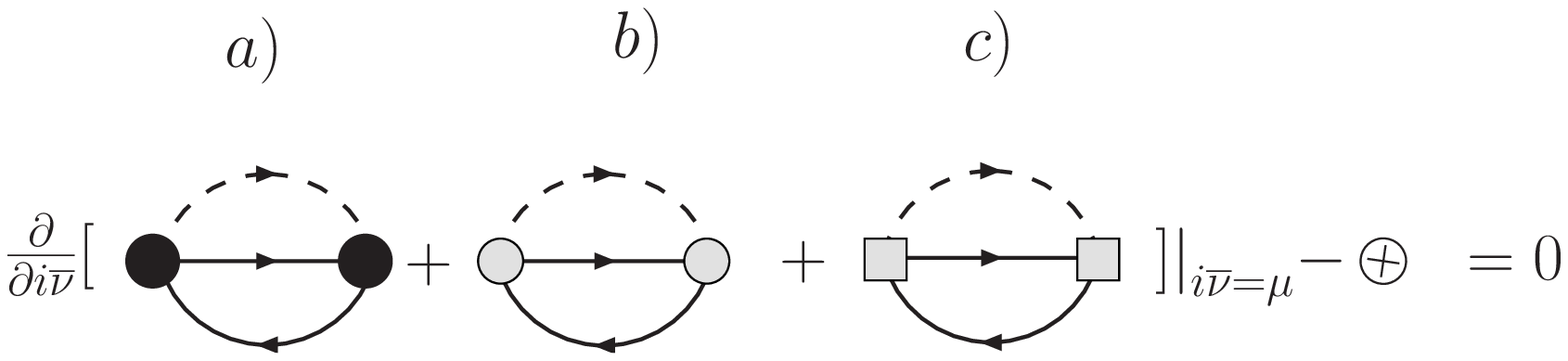}
\end{center}
\caption{
Recursion relation for the impurity field renormalization factor $(Z_b-1)$ (crossed dot).
}
\label{fig:bosonself}
\end{figure}

The remaining diagram $(k)$ evaluates to
\begin{eqnarray}
(k)&=& \mu^{2\epsilon} \int_0^\infty d x d y  \frac{x^\epsilon y^\epsilon}{(x+y+1)(x+y)}  \nonumber \\
&=& -\mu^{2\epsilon} \left( \frac{1}{2\epsilon}+\mathcal{O}(\epsilon^0)\right).
\end{eqnarray}
We will re-encounter the same
integral within the calculation of the field renormalization factor.
In order to arrive at the final two loop $Z$-factors we dress the integral values with the appropriate interaction vertices. After some calculation we find
\begin{eqnarray}
Z_\alpha&=& 1-\frac{2}{\epsilon} \left(\beta+\gamma-\frac{\beta \gamma}{2} \right)\nonumber \\
&+&\frac{2}{\epsilon^2}\left(\alpha^2+2\left(\beta^2+\beta \gamma+\gamma^2 \right) \right) \; ,  \nonumber \\
Z_\beta &=&1-\frac{2}{\epsilon} \left(\frac{\alpha^2}{2\beta}+\beta-\frac{\beta^2}{4}
-\frac{\alpha^2 \gamma}{4 \beta}-\frac{\gamma^2}{4} \right) \nonumber \\
&+&\frac{2}{\epsilon^2} \left(2 \left(\beta^2+\alpha^2 \right)
+\frac{\alpha^2 \gamma}{\beta} \right) \; , \nonumber \\
Z_\gamma &=& 1-\frac{2}{\epsilon} \left (\frac{\alpha^2}{2 \gamma}+\gamma-\frac{\gamma^2}{4}
-\frac{\alpha^2 \beta}{4 \gamma}-\frac{\beta^2}{4} \right) \nonumber \\
&+& \frac{2}{\epsilon^2} \left(2 \left(\gamma^2+\alpha^2 \right)
+\frac{\alpha^2 \beta}{\gamma} \right) \; .
\end{eqnarray}
The next step consists of the propagator renormalization, whose recursion relation is shown in Fig.~\ref{fig:bosonself}. We calculate the diagram shown in Fig.~\ref{fig:bosonself} (a). The dependence on the external frequency can be extracted by calculating
\begin{eqnarray}
&&\Sigma (i \nu_n-\lambda)-\Sigma (0) \nonumber \\ &=& (i\nu_n-\lambda) \mu^{2\epsilon} \int_0^{\infty} d x d y \frac{x^\epsilon y^\epsilon}{(x+y+1)(x+y)} \nonumber \\ &=& -(i\nu_n-\lambda)\mu^{2\epsilon} \left(\frac{1}{2\epsilon}+\mathcal{O}(\epsilon^0) \right)\; .
\end{eqnarray}

If we dress this integral with the appropriate vertices and apply the renormalization
condition shown in Fig.~\ref{fig:bosonself} (note that $i\overline{\nu}_n=i\nu_n-\lambda$) we find
\begin{eqnarray}
Z_b=1+\frac{1}{2\epsilon}(\alpha^2+\beta^2+\gamma^2) \; .
\end{eqnarray}
In order to derive the RG beta functions we use the fact that the bare couplings have no
knowledge about the renormalization condition, which can be expressed as
\begin{eqnarray}
 \mu \frac{d\alpha_0}{d\mu}=\mu \frac{d\beta_0}{d\mu}
=\mu \frac{d\gamma_0}{d\mu}=0 \; .
\end{eqnarray}
This leads to a set of coupled equations
\begin{widetext}
\begin{eqnarray}
&-&\epsilon \alpha+\alpha \frac{\partial \ln
\left(\frac{Z_\alpha}{Z_f} \right)}{\partial \alpha} \beta(\alpha)
+\alpha \frac{\partial \ln \left(\frac{Z_\alpha}{Z_f} \right)}{\partial \beta}
\beta(\beta)+\alpha \frac{\partial \ln \left(\frac{Z_\alpha}{Z_f} \right)}{\partial \gamma}
\beta(\gamma)=0 \; , \nonumber \\
&-&\epsilon \beta+\beta \frac{\partial \ln \left(\frac{Z_\beta}{Z_f} \right)}{\partial \alpha}
\beta(\alpha)+\beta \frac{\partial \ln
\left(\frac{Z_\beta}{Z_f} \right)}{\partial \beta} \beta(\beta)+\beta \frac{\partial \ln \left(\frac{Z_\beta}{Z_f} \right)}{\partial \gamma}
\beta(\gamma)=0 \; , \nonumber \\
&-&\epsilon \gamma+\gamma \frac{\partial
\ln \left(\frac{Z_\gamma}{Z_f} \right)}{\partial \alpha} \beta(\alpha)
+\gamma \frac{\partial \ln \left(\frac{Z_\gamma}{Z_f} \right)}{\partial \beta}
\beta(\beta) +\gamma \frac{\partial \ln \left(\frac{Z_\gamma}{Z_f} \right)}{\partial \gamma}
\beta(\gamma)=0 \; ,
\end{eqnarray}
\end{widetext}
where the introduced $\beta$ functions correspond to
$\beta (\alpha) = \mu \frac{d}{d \mu}\alpha$ etc.
These coupled equations can be solved iteratively,
leading to
\begin{eqnarray}
\beta(\alpha)&=&\epsilon \alpha + \alpha \left(\alpha^2+\left(\beta-\gamma \right)^2 \right)
+2\alpha(\beta+\gamma) \; , \nonumber \\
\beta (\beta) &=& \epsilon \beta +2 \beta \left (\frac{\alpha^2}{2\beta}+\beta \right)
+\beta \left(\alpha^2-\frac{\alpha^2 \gamma}{\beta} \right) \; ,\nonumber \\
\beta (\gamma) &=&\epsilon \gamma+ \left(\alpha^2+2\gamma^2 \right)
-\alpha^2 (\beta-\gamma) \; .
\end{eqnarray}
In order to arrive at the actual bosonic Kondo model (Eq.~\eqref{equ:bosonickondomodel})
with potential scattering we have to make the following identifications
\begin{eqnarray}
\alpha=\frac{j_\perp}{2}, \quad \beta=\frac{{j_z}}{4}+\frac{v}{4}\; , \quad \textrm{and}
\quad \gamma=-\frac{j_z}{4}+\frac{v}{4} \; .
\label{parameterchoice}
\end{eqnarray}
The flow equations of the physical couplings $j_\perp,j_z,v$ are related to the
intermediate quantities $\alpha,\beta,\gamma$ according to
\begin{eqnarray}
\beta(j_\perp)&=&2 \beta (\alpha) \; , \nonumber \\
\beta(j_z)&=&2 \left[ \beta(\beta)-\beta(\gamma) \right] \; ,\nonumber \\
\beta(v)&=&2 \left[ \beta(\beta)+\beta(\gamma) \right] \; ,
\end{eqnarray}
with the interaction parameters chosen as shown in Eq.~\eqref{parameterchoice}.
This leads to the following flow equations
\begin{eqnarray}
\beta (j_\perp)&=&\epsilon j_\perp+vj_\perp+\frac{j_\perp j_z^2}{4}+\frac{j_\perp^3}{4} \; , \nonumber \\
\beta (j_z) &=& \epsilon j_z+v j_z +\frac{j_\perp^2 j_z}{4} \; ,\nonumber \\
\beta (v) &=& \epsilon v +\frac{v^2}{2}+j_\perp^2+\frac{j_z^2}{2},
\end{eqnarray}
or in the spin-isotropic situation
\begin{eqnarray}
\beta (j)&=&\epsilon j+vj+\frac{j^3}{2} \; , \nonumber \\
\beta (v)&=&\epsilon v+\frac{v^2}{2}+\frac{3}{2} j^2
\end{eqnarray}
which are the RG equations \eqref{RGeq} quoted in the main text.

The RG equations for the model of canonical bosons, Eq.~\eqref{canRGeq} in Sec.~\ref{canonical},
are obtained from the same diagrammatic expansion
(albeit with different numerical values for the diagrams).
The observation that the RG equations are exact to {\it all} orders comes
can be easily understood by noting that
the counter\-terms cancel all higher-order contributions
because the only non-vanishing diagrams are ladder-like (in electronic language one would
call these diagrams the Cooper diagrams), see Ref.~\onlinecite{Sachdevbook}.

\section{RG for the bosonic Anderson model}
\label{appen:bosonicandersonrg}

\begin{figure}[h]
\includegraphics[width=0.4\textwidth]{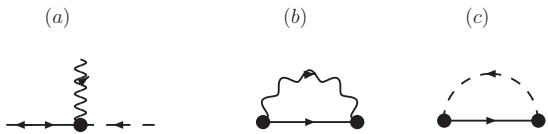}
\caption{
(a) Interaction vertex between bulk and impurity in the bosonic Anderson model;
the wiggly line denotes the
particle $b_0$, the dashed line is $b_\sigma$ and the full line stands for $z_\sigma$
(incoming arrow) or $z^*_\sigma$ (outgoing arrow).
(b) and (c) denote self-energy corrections to the local levels $b_\sigma$ (b) and $b_0$ (c).
}
\label{fig:vertexdef}
\end{figure}

Within this section we derive the RG equations for the bosonic version of the
infinite-$U$ Anderson model of Sec.~\ref{sec:boseanderson}.
After a Fourier transformation for the quadratic bulk part,
the action of the model \eqref{bosoAnd} reads
\begin{eqnarray}
\mathcal{S}&=&\frac{1}{\gz}\beta^{-1}\sum_{\nu_n,\sigma}\int
\frac{d^d{k}}{(2 \pi)^d} z_\sigma^* (\nu_n,k)
\left[ \nu_n^2+k^2 \right] z_\sigma(\nu_n,k)\nonumber \\
&+& u_0 \int d^d x d\tau z^*_\sigma z_\sigma z^*_{\sigma'}z_{\sigma'} \nonumber \\
&+&\int d\tau \overline{b}_0(\partial_\tau+\lambda)b_0
+\int d\tau \overline{b}_\sigma (\partial_\tau+\lambda)b_\sigma \nonumber \\
&+& w_0 \int d\tau \left[
\overline{b}_\sigma b_0 z_\sigma (\tau) +\textrm{h.c.} \right].
\end{eqnarray}
The flow equations for the bulk interaction $u_0$ and the hybridization $w_0$ will not mix to
one-loop order.
In the following we will ignore the renormalization of the
mass difference between the different impurity levels [$\epsilon_0$ in \eqref{bosoAnd}],
i.e., we work at criticality.
We introduce renormalized couplings according to
\begin{eqnarray}
w_0 &=& \mu^{\epsilon}\sqrt{\frac{2}{\gz S_d}}\frac{Z_w}{\sqrt{Z_{b_0}Z_{b_\sigma}Z_z}}w \quad \textrm{and}\nonumber \\
u_0 &=& \mu^{2\epsilon}\frac{4}{S_d \gz^2}\frac{Z_u}{Z_z^2} u \; ,
\end{eqnarray}
where $\epsilon=\frac{3-d}{2}$.
It is interesting to note that technically we
have a version of the RG of a model which is at its upper critical dimension (both for the bulk
and the impurity), whereas the RG for the bosonic Kondo model is reminiscent of an RG at
the lower critical dimension.
The vertices of the theory are shown in Fig.~\ref{fig:vertexdef}.

The calculation for the renormalization of the hybridization $w$ completely parallels the
analysis in the pseudogap infinite-$U$ Anderson model, which was presented in
Ref.~\onlinecite{Fritz}, and therefore without showing further calculational details leads to
the same flow equation, namely
\begin{eqnarray}
\frac{d w}{d\ln\mu}=\epsilon w-\frac{3}{2}w^3.
\end{eqnarray}
In addition, we have to calculate the perturbative corrections to the bulk interaction
vertex, which is identical to the usual O(4)-$\phi^4$ theory.
In our convention the result is:
\begin{eqnarray}
\frac{du}{d\ln\mu} = -2 \epsilon u+12 u^2 \; .
\end{eqnarray}

The hidden O(4) symmetry of the theory becomes apparent upon writing the bosonic spinons
as a 4-vector $(\mathfrak{Re}
z_\uparrow,\mathfrak{Im}z_\uparrow,\mathfrak{Re}z_\downarrow,\mathfrak{Im}z_\downarrow)$.

\section{Fermionic Kondo problem at the band edge}
\label{appen:fermions}

Remarkably, the two-level system coupled to the two-component bosonic gas
studied in Sec.~\ref{canonical} turns out to follow a weak-coupling
RG flow indentical to that of a fermionic Kondo problem,
where the chemical potential is tuned to the (e.g. lower) band edge
(i.e. with full particle-hole asymmetry).
The equivalence can best be understood in terms of the local density of states --
the only quantity which determines the impurity behavior --
which in both cases (canonical bosons and fermions)
is of the form $\rho (\omega)=\rho_0 |\omega|^r \theta (\omega)$.

The strict particle-hole asymmetry of the electrons implies that there are only particle
excitations. Introducing dimensionless couplings according to standard
practice\cite{Fritz} we can derive the following flow equations
for the fermionic problem:
\begin{eqnarray}\label{eq:asymm}
\beta (j)&=& rj-\frac{j^2}{2}+\frac{vj}{2}\; ,\nonumber \\
\beta(v)&=&rv+\frac{3}{4}j^2+\frac{1}{4}v^2 \; ,
\end{eqnarray}
These equations are contained in the earlier analysis of Ref.~\onlinecite{Zawadowski},
and are identical to those obtained for the problem of canonical bosons in
Sec.~\ref{canonical}. Again, the equations are exact to all orders in
perturbation theory (!).

For a general fermionic metallic ($r=0$) Kondo problem with band asymmetry
(i.e. the chemical potential not being pinned to the center of the band)
one has to perform a two-step RG procedure:
Initially, one has to follow the fully asymmetric scaling, Eq.\eqref{eq:asymm}.
Once the remainder of the band is symmetric,
the usual poor man's scaling equations take over --
here the potential scattering term is a marginal operator.

We note that the fully asymmetric fermionic Kondo problem can be relevant
in the context of DMFT for Hubbard models:
The transition from the undoped to the doped Mott insulator
upon variation of the chemical potential happens precisely
when the chemical potential reaches the gap edge
(provided the transition is continuous).



\begin{thebibliography}{99}

\bibitem{Sachdevbook} S. Sachdev, {\it Quantum Phase Transitions},
Cambridge University Press, Cambridge (1999).

\bibitem{Metzner} W. Metzner and D. Vollhardt, Phys. Rev. Lett. {\bf 62}, 324 (1989).

\bibitem{Georges} A. Georges, G. Kotliar, W. Krauth, and M. J. Rozenberg, Rev.
Mod. Phys. {\bf 68}, 13 (1996).

\bibitem{Withoff} D. Withoff and E. Fradkin, Phys. Rev. Lett. {\bf 64}, 1835 (1990).

\bibitem{Hewson} A. C. Hewson, {\it The Kondo Problem to Heavy
Fermions}, Cambridge University Press, Cambridge (1996).

\bibitem{GBI}
C.~Gonzalez-Buxton and K.~Ingersent, \prb {\bf 57}, 14254 (1998).

\bibitem{Fritz} L. Fritz and M. Vojta, Phys. Rev. B {\bf 70}, 214427 (2004).

\bibitem{Fritz1} L. Fritz, S. Florens, and M. Vojta, Phys. Rev. B {\bf 74}, 144410 (2006).

\bibitem{Cox} D. L. Cox and A. Zawadowski, Adv. Phys. {\bf 47}, 599 (1998).

\bibitem{Potok} R. M. Potok, I. G. Rau, H. Shtrikman, Y. Oreg, and D.
Goldhaber-Gordon, cond-mat/0610721.

\bibitem{Jones} B. A. Jones, C. M. Varma, and J. W. Wilkins,
Phys. Rev. Lett. {\bf 61}, 125 (1988);
B. A. Jones and C. M. Varma,  Phys. Rev. B {\bf 40}, 324 (1989).

\bibitem{ALJ}
I. Affleck, A. W. W. Ludwig, and B. A. Jones, Phys. Rev. B {\bf 52}, 9528 (1995).

\bibitem{Zarand} G. Zar\'{a}nd, C.-H. Chung, P. Simon, and M. Vojta, Phys. Rev.
Lett. {\bf 97}, 166802 (2006).

\bibitem{Matthiasreview} M. Vojta, Phil. Mag. {\bf 86}, 1807 (2006).

\bibitem{Zwerger} A. J. Leggett, S. Chakravarty, A. T. Dorsey, M. P. A. Fisher,
A. Garg, and W. Zwerger, Rev. Mod. Phys. {\bf 59}, 1 (1987).

\bibitem{VTB}
M. Vojta, N.-H. Tong, and R. Bulla,
\prl {\bf 94}, 070604 (2005).

\bibitem{vojta0}
S. Sachdev, C. Buragohain, and M. Vojta,
Science {\bf 286}, 2479 (1999).

\bibitem{vojta}
M. Vojta, C. Buragohain, and S. Sachdev,
Phys. Rev. B {\bf 61}, 15152 (2000).

\bibitem{ssmv03}
S. Sachdev and M. Vojta, \prb {\bf 68}, 064419 (2003).

\bibitem{troyer}
M. Troyer, Prog. Theor. Phys. Supp. {\bf 145}, 326 (2002).

\bibitem{sandvik}
K. H. H\"oglund and A. W. Sandvik,
\prl {\bf 91}, 077204 (2003) and
cond-mat/0701472.

\bibitem{Si} J. L. Smith and Q. Si, Europhys. Lett. {\bf 45}, 228 (1999).

\bibitem{Sengupta} A. M. Sengupta, Phys. Rev. B {\bf 61}, 4041 (2000).

\bibitem{Zhu} L. Zhu and Q. Si, Phys. Rev. B {\bf 66}, 024426 (2002).

\bibitem{Zarand2} G. Zar\'{a}nd and E. Demler, Phys. Rev. B {\bf 66}, 024427 (2002).

\bibitem{MVMK}
M. Vojta and M. Kir\'{c}an, \prl {\bf 90}, 157203 (2003).

\bibitem{LeHur} K. Le Hur, Phys. Rev. Lett. {\bf 92}, 196804 (2004) .


\bibitem{Florens} S. Florens, L. Fritz, and M. Vojta,
Phys. Rev. Lett. {\bf 96}, 036601 (2006).

\bibitem{Park} S. Sachdev and K. Park, Annals of Physics (N.Y.) {\bf 298}, 58 (2002).

\bibitem{Zoller} A. Recati, P. O. Fedichev, W. Zwerger, J. von Delft
 and P. Zoller, Phys. Rev. Lett. {\bf 94}, 040404 (2005).

\bibitem{williams} T. Nikuni and J. E. Williams,
J. Low Temp. Phys. {\bf 133}, 516 (2003).

\bibitem{chubukov} A. V. Chubukov, T. Senthil, and S.
Sachdev, Phys. Rev. Lett. {\bf 72}, 2089 (1994).

\bibitem{isakov} S. V. Isakov, T. Senthil, and Y. B. Kim, Phys. Rev. B {\bf 72}, 174417 (2005).

\bibitem{azaria} P. Azaria, P. Lecheminant, and D. Mouhanna, Nucl. Phys. B {\bf 455}, 648 (1995).

\bibitem{ssmorinari}
S. Sachdev and T. Morinari, Phys. Rev. B {\bf 66}, 235117 (2002).


\bibitem{Auerbach} A. Auerbach, {\it Interacting Electrons and Quantum Magnetism},
Springer (1994).

\bibitem{Fradkin} E. Fradkin, {\it Field Theories of Condensed Matter Systems},
Addison-Wesley Publishing Company (1991).

\bibitem{Shenker} E. Fradkin and S. H. Shenker,
Phys. Rev. D {\bf 19}, 3682 (1979).

\bibitem{ChubuStar} A. V. Chubukov and O. A. Starykh, Phys. Rev. B, {\bf 52},
440 (1995).

\bibitem{deconfined} T. Senthil, L. Balents, S. Sachdev, A. Vishwanath, and M.
P. A. Fisher, Phys. Rev. B {\bf 70}, 144407 (2004).

\bibitem{Motrunich} O. I. Motrunich and A. Vishwanath, Phys. Rev. B {\bf 70}, 075104 (2004).

\bibitem{Kolezhuk} A. Kolezhuk, S. Sachdev, R. R. Biswas, and P. Chen, Phys.
Rev. B {\bf 74}, 165114 (2006).

\bibitem{Zinn} J. Zinn-Justin, {\it Quantum Field Theory and Critical
Phenomena}, Oxford University Press (1996).

\bibitem{barvsstar}
Note that our notations for conjugate objects, $\overline{b}$ and $z^*$,
aim at distinguishing canonical bosons $b$ from commuting numbers $z$.
Of course, in the language of coherent-state path integrals the difference is
in the dynamic term.

\bibitem{lambda}
A. Zawadowski and P. Fazekas, Z. Physik {\bf 226}, 235 (1969);
P. Coleman, Phys. Rev. B {\bf 29}, 3035 (1984).

\bibitem{costi}
T. A. Costi, J. Kroha, and P. W\"olfle,
\prb {\bf 53}, 1850 (1996).

\bibitem{lee}
H.-J. Lee and R. Bulla, cond-mat/0606325.

\bibitem{bergmann}O. Bergman, Phys. Rev. D {\bf 46}, 5474 (1992).

\bibitem{Zawadowski}
O. Ujsaghy, K. Vladar, G. Zarand, and A. Zawadowski,
J. Low Temp. Phys. {\bf 126}, 1221 (2002).

\bibitem{coldea0}
R. Coldea, D. A. Tennant, A. M. Tsvelik, and Z. Tylczynski,
\prl {\bf 86}, 1335 (2001).

\bibitem{coldea}
R. Coldea, D. A. Tennant, and Z. Tylczynski,
Phys. Rev. B {\bf 68}, 134424 (2003).

\bibitem{Jaxodraw}
D. Binosi and L. Theussl, hep-ph/0309015 and http://jaxodraw.sourceforge.net.

\end{thebibliography}
\end{document}